# Thermal decomposition of norbornane (bicyclo[2.2.1]heptane) dissolved in benzene. Experimental study and mechanism investigation.

*Olivier HERBINET,\* Baptiste SIRJEAN, Frédérique BATIN-LECLERC, René FOURNET and Paul-Marie MARQUAIRE*

Département de Chimie Physique des Réactions, UMR 7630 CNRS, Nancy Université - ENSIC, 1 rue Grandville, 54001, Nancy, France.

CORRESPONDING AUTHOR FOOTNOTE

\*E-mail: herbinet@ensic.inpl-nancy.fr, Fax: +333 83 37 81 20; Tel: +333 83 17 50 06

ABSTRACT

The thermal decomposition of norbornane (dissolved in benzene) has been studied in a jet stirred reactor at temperatures between 873 and 973 K, at residence times ranging from 1 to 4 s and at atmospheric pressure, leading to conversions from 0.04 to 22.6%. 25 reaction products were identified and quantified by gas chromatography, amongst which the main ones are hydrogen, ethylene and 1,3-cyclopentadiene. A mechanism investigation of the thermal decomposition of the norbornane – benzene binary mixture has been performed. Reactions involved in the mechanism have been reviewed: unimolecular initiations by C–C bond scission of norbornane, fate of the generated diradicals, reactions of transfer and propagation of norbornyl radicals, reactions of benzene and cross–coupling reactions.

KEYWORDS

Pyrolysis, Norbornane, Bicyclo[2.2.1]heptane, Mechanism investigation, Low conversion, Diradical mechanism.

MANUSCRIPT TEXT

Introduction

While the thermal decomposition of cyclanes has been the subject of several papers, there are only few studies about the reactions of polycyclanes and the corresponding kinetic parameters are still very uncertain.

The geometry and the enthalpy of formation of norbornane or bicyclo[2.2.1]heptane ($C_7H_{12}$), a bridged bicyclic alkane (Figure 1), have been previously studied in order to relate the structure of the molecule to its strain.[1,2] A ring strain energy of 17,2 kcal.mol$^{-1}$ has been estimated for norbornane[2] (the ring strain energy is defined as the difference between the experimental gas phase enthalpy of formation and the gas phase enthalpy estimated using the group additivity method proposed by Benson[3] for the estimation of thermochemical data). Baldwin and al.[4] studied the thermal isomerization of 3-butenyl-cyclopropane to norbornane. It was observed that 3-butenyl-cyclopropane lead to norbornane and many other hydrocarbons when heated at the temperature of 688 K. It was also established that the formation of norbornane occurs through the scission of the C2–C3 bond of 3-butenyl-cyclopropane and might involve the initial generation of a diradical (Figure 2).

**Figure 1**

**Figure 2**

O'Neal and Benson studied the kinetics of pyrolysis of some non-bridged bicyclic alkanes (e.g. bicyclo[2.2.0]hexane, bicyclo[3.2.0]heptane) from the point of view of diradical intermediates.[5] Diradical mechanism estimates were found to be consistent with experimental Arrhenius parameters for a large number of hydrocarbons. Reaction channels for the fate of diradicals were proposed by Tsang in the case of cyclopentane and cyclohexane.[6,7] Direct studies of trimethylene and tetramethylene diradicals have been performed by Pedersen and al. by using femtosecond laser techniques with mass spectrometry in a molecular beam.[8] More recently, Sirjean et al. performed quantum calculations about the gas phase unimolecular decomposition of cyclobutane, cyclopentane and cyclohexane by using a diradical mechanism.[9] The theoretical approach used for the calculation was validated by comparing calculated results with available experimental data. Several papers about the pyrolysis of tricyclo[5.2.1.0$^{2,6}$]decane, a tricyclic alkane, have been published.[10-14] Indeed this hydrocarbon is a component of synthetic fuels used in aeronautics. A comprehensive primary mechanism of the pyrolysis of this species has been developed in our laboratory.[14] The reactions of unimolecular initiation of this polycyclic compound have been detailed and the reactions of diradicals (decompositions by β-scission, internal disproportionnations) have been taken in account on a systematic way.

The first purpose of this article is to present new experimental results of the pyrolysis of norbornane (solid at room temperature) dissolved in benzene. In line with previous work on hydrocarbons,[14-17] experiments have been performed in a jet stirred reactor which was operated at temperatures between 873 and 973 K, residence times between 1 and 4 s, at a pressure of 106 kPa and at high dilution. Conversions ranging from 0.04 to 22.6% have been obtained. The attention has been paid to the analysis of the products of the reaction. The formation of 25 major and minor products has been observed. The second objective of this paper is to describe the reactions involved in the mechanism of the pyrolysis of the norbornane – benzene binary mixture. These reactions include all the possible channels of unimolecular initiation of norbornane using a diradical approach as in the case of tricyclo[5.2.1.0$^{2,6}$]decane.[14] Cross-coupling reactions due to the presence of benzene in the feed of the reactor have been also reviewed although it will be shown later in this paper that benzene is not very reactive under the operating conditions of this study.

Experimental Section

The apparatus (Figure 3) used for the experimental study of the thermal decomposition of norbornane (dissolved in benzene) has already been described in two papers about the pyrolysis of tricyclo[5.2.1.0$^{2,6}$]decane[14] and n-dodecane,[15] respectively Its main features are reminded and specificities linked to the dissolution of norbornane in benzene are discussed below.

**Figure 3**

Experiments were performed in a continuous quartz jet stirred reactor operated at constant temperature (inner volume of about 90 cm$^3$). This reactor was designed to be perfectly stirred for residence times ranging from 0.5 to 5 s.[18,19] The heating of the reactor was achieved using Thermocoax heating resistors coiled around the vessel. Temperature inside the reactor was measured with a type K thermocouple which was located at the level of the injection cross at the center of the vessel. Before entering the reactor, reactants were preheated to the reaction temperature to avoid the formation of temperature gradients inside the gas phase due to the endothermal properties of the pyrolysis reaction. The residence time of the reactants inside the annular preheater was very short, i.e. about 1% of the residence time inside the reactor. Pressure inside the reactor was set equal to 106 kPa and was controlled with a control valve set downstream the products analysis devices.

Unlike most hydrocarbons having close molecular weights, norbornane (C$_7$H$_{12}$) is solid at room temperature. The melting point of pure norbornane at atmospheric pressure is 360 K[20] and its boiling point is 381 K.[21] In order to study the pyrolysis of norbornane with the same apparatus as that used for n-dodecane and tricyclo[5.2.1.0$^{2,6}$]decane,[14,15] solid norbornane has been dissolved in a solvent. Benzene was chosen since it is a good solvent for many hydrocarbons and because, as an aromatic compound, it is very unreactive at low temperature. Norbornane used for the experiments was provided by Aldrich (mass fraction purity greater than 0.98) and benzene was provided by Fluka (mass fraction purity greater than 0.99).

The liquid reactant (20 wt% norbornane, 80 wt% benzene) was stored in a pressurized glass vessel. Before performing experiments nitrogen bubbling and vacuum pumping were performed in order to remove oxygen traces dissolved in the hydrocarbon mixture. The liquid reactant mass flow rate was controlled by a mass flow controller, mixed to the carrier gas (helium 99.995% pure) and evaporated in a single pass heat exchanger, the temperature of which was set above the boiling point of the diluted hydrocarbon mixture. The molar composition of the mixture at the inlet of the reactor was 0.7% norbornane, 3.6% benzene and 95.7% helium.

Products leaving the reactor have been analyzed by gas chromatography. Analyses were performed in two steps. Light species (which are gaseous at room temperature such as hydrogen and hydrocarbons containing less than 5 carbon atoms) were analyzed on-line by two gas chromatographs. The first chromatograph was fitted with a carbosphere packed column and both a flame ionization detector (FID) for the detection of methane and C2 hydrocarbons and a thermal conductivity detector (TCD) for the detection of hydrogen. Argon was chosen as carrier and reference gas in order to detect hydrogen with a better sensibility. A first analysis was performed with a constant oven temperature of 303 K to separate the hydrogen peak from the helium peak (experimental carrier gas). A second analysis was performed with a constant oven temperature of 473 K for the hydrocarbons separation. Retention times (in min) were: methane: 2.4, acetylene: 5.3, ethylene: 7.3 and ethane: 9.6. The second chromatograph used for light species analyses was equipped with a FID for the hydrocarbons detection and a Haysep D packed column. *This column gave a good separation for hydrocarbons from methane to $C_5$ hydrocarbons. In particular the peaks corresponding to species like allene and propene or like 1-butene, 2-butene, 1,3-butadiene and 1-butyne were well defined. Retention times (in min) for species whose formation was observed during the study were: methane: 2.6, ethane: 15.5, propene: 61.2, allene: 70.9, propyne: 73.5, 1-butene: 106.6, 1,3-butadiene: 107.7 and 1,3-cyclopentadiene: 147.8.* Species identification and calibration were performed with gaseous standard mixtures provided by Air Liquide and Messer. Heavy species (hydrocarbons containing more than 5 carbon atoms which are liquid or solid at room temperature) were condensed in a trap connected at the outlet of the reactor and maintained at liquid nitrogen temperature during a determined period of time. After this time of accumulation, the trap was disconnected and solvent (acetone) and a known amount of internal standard (n-octane) were added. When the temperature of the trap was back to a temperature close to 273 K the mixture was poured into a sampling bottle and then analyzed by gas chromatography. *A first analysis was performed with a gas chromatograph fitted with a capillary HP-1 column and a FID for the separation and the detection of hydrocarbons. Oven temperature profile was set to: 313 K held 30 min, rate 5 K.min$^{-1}$, 453 K held 62 min in order to obtain a good separation of the products of the reaction. Retention times of main products of the reaction (in min) were: 1,3-cyclopentadiene: 3.7, benzene: 4.9, norbornane: 7.3, toluene: 7.9, styrene: 17.4, indene: 38.8, naphthalene: 46.3, biphenyl: 53.3. Calibration was performed with prepared solutions containing small amounts of quantified hydrocarbons and of n-octane (internal standard). A second analysis was performed for the identification of the products of the reaction with a gas chromatography-mass spectrometry system working in the same conditions than the gas chromatograph used for quantification (same column, same carrier gas, same carrier gas flow rate, same oven temperature profile). This procedure allowed us to obtain the same chromatograms with both chromatographs so that direct comparison of the peaks could be performed. Identification of the products separated by the HP-1 column was performed by comparison of the mass spectrum corresponding to the detected peaks with the numerous mass spectra included in the library NBS 75K which was provided by Agilent with the GC-MS apparatus.*

*The consistency between the different chromatographic analyses was verified from products which were present on two chromatograms (like 1,3-cyclopentadiene, methane and ethane). In each case the relative variation between the mole fractions corresponding to the different analyses was less than 5%.* The repeatability of experimental results has been studied. Calculated maximum uncertainties in the experimental mole fractions were ±5% for species analyzed on-line and ±8% for heavy species condensed in the trap. *Carbon to hydrogen ratios (C/H ratios) in the products have been calculated. For*

*this calculation all species have been taken in account except norbornane (this species has no influence on the value of the C/H ratio), benzene, toluene, styrene, naphthalene and biphenyl (these four last species mainly come from benzene). An average value of 0.60 (± 0.02) has been obtained. This value is slightly above the theoretical value (7/12=0.58) but it should be kept in mind that a rigorous distinction between products from the norbornane and from the benzene (C/H ratio of 1) is not possible.*

Experimental Results

**Norbornane – benzene binary mixture.** The evolution of the conversion of norbornane with residence time is shown on Figure 4. Because the values of the difference between the mass flow rates of norbornane entering and leaving the reactor were not accurate enough for such low conversions, the values of conversion presented on this graph were deduced from the products of the reaction apart from aromatic and polyaromatic compounds (toluene, styrene, indene, naphthalene and biphenyl produced in very small quantities). Under the conditions of the study, these last products probably derived from benzene. At higher conversions aromatic and polyaromatic compounds could be formed through secondary reactions from small unsaturated hydrocarbons and from 1,3-cyclopentadiene. No evolution of the mass flow rate of benzene was observed between the inlet and the outlet of the reactor under the operating conditions of our study. Benzene appeared to be very stable corresponding to very low conversions.

**Figure 4**

Twenty five products of the thermal decomposition of the norbornane – benzene binary mixture have been analyzed. These products are (by increasing molecular weight): hydrogen, methane, acetylene, ethylene, ethane, allene, propyne, propene, 1-butene, 1,3-butadiene, 1,3-cyclopentadiene, 1,3-cyclohexadiene, 1,4-cyclohexadiene, 5-methyl-1,3-cyclopentadiene, 1,3,5-hexatriene, toluene, 3-ethyl-cyclopentene, ethenyl-cyclopentane, 4-methyl-cyclohexene, methylene-cyclohexane, styrene, indene, naphthalene and biphenyl. An unidentified minor product (molecular weight of 94 g.mol$^{-1}$ according to mass spectroscopy) has been detected between toluene and styrene.

It is worth noticing that the formation in small amounts of several species having the same molecular weight as norbornane has been observed: 3-ethyl-cyclopentene, ethenyl-cyclopentane, 4-methyl-cyclohexene and methylene-cyclohexane (Figure 5). The evolution of the mole fractions of these four products with residence time is shown on Figure 6. The possible channels of formation of these particular species which were observed even at very low conversion (less than 0.5%) will be discussed later in this paper. The formation of very small quantities of aromatic and polyaromatic compounds such as toluene, styrene, indene, naphthalene and biphenyl was also observed. These species probably come from reactions of benzene or from cross-coupling reactions of the norbornane – benzene binary mixture. The presence of benzene in the feed of the reactor masked the possible formation of small quantities of this species as specific product from norbornane.

**Figure 5**

**Figure 6**

Figure 7 displays the distribution of the products in term of selectivity (here the selectivity of a product is defined as the ratio of the mole fraction of the considered product and the sum of the mole fractions of all products) at a temperature of 973 K and a residence time of 1 s. This figure shows that the three main products of the reaction are hydrogen, ethylene and 1,3-cyclopentadiene which are formed in similar quantities. Figure 8 shows the evolution of the mole fraction with residence time of these three main products, as well as methane, propene, 1,3-butadiene, toluene and biphenyl.

**Figure 7**

**Figure 8**

The primary products of the reaction of the thermal decomposition of norbornane dissolved in benzene were determined from a study of the selectivity performed at a temperature of 953 K (corresponding to a maximum conversion of 15%). A species is probably a primary product if the extrapolation to origin of its selectivity versus residence time gives a value different from zero (corresponding to a non zero initial rate of production). According to this study 15 species seem to be primary products: hydrogen, methane, ethylene, ethane, propene, 1-butene, 1,3-butadiene, 1,3-cyclopentadiene, 1,3-cyclohexadiene, toluene, 3-ethyl-cyclopentene, 4-methyl-cyclohexene, methylene-cyclohexane, ethenyl-cyclopentane and biphenyl. The values of selectivities to origin of these products are given in Table 1. Species with the highest selectivities at origin are hydrogen (0.346), ethylene (0.314) and 1,3-cyclopentadiene (0.184). Unlike toluene and biphenyl (Figure 9a) other aromatic and polyaromatic compounds (styrene, indene and naphthalene) do not seem to be primary products (extrapolations to origin of their selectivities versus residence time are close to zero; Figure 9b). This can be explained by the fact that toluene and biphenyl can be obtained directly from phenyl radicals derived from benzene (by combination of phenyl radical and methyl radical for toluene, by self-combination of phenyl radicals or by ipso addition of phenyl on benzene for biphenyl) whereas styrene, indene and naphthalene can not be directly formed from primary radicals generated by the decomposition of norbornane and benzene. Amongst species having the same molecular weight as norbornane 3-ethyl-cyclopentene has the largest selectivity to origin.

**Table 1**

**Figure 9**

**Influence of the benzene on the kinetic of the reaction.** A short study of the thermal decomposition of pure benzene was performed in order to determine if benzene plays a role in the kinetics of the reaction of pyrolysis of the norbornane – benzene binary mixture. Experiments were performed at temperatures between 913 and 973 K, at residence times ranging from 1 to 4 s and at a pressure of 106 kPa. The molar composition of the flow entering the reactor was 96.4% helium and 3.6% benzene (the mole fraction of benzene at the inlet of the reactor was fixed to the same value as in the case of the study of the binary mixture norbornane – benzene for direct comparison).

Figure 10 shows the evolution of the conversion of benzene with residence time at temperatures between 913 and 973 K (the conversion were deduced from the products of the reaction). Under these operating conditions benzene appeared to be very stable. A maximum conversion of benzene of $8 \times 10^{-2}$

% was obtained at a temperature of 973 K and a residence time of 1 s. For comparison the conversion of norbornane (dissolved in benzene) was 9.7% under the same conditions.

**Figure 10**

The only product of the reaction which was detected was biphenyl. Hydrogen is probably another product of the reaction[22-25] but it was not detected (TCD is known for being a much less sensitive detector than the FID). The formation of toluene, styrene, indene and naphthalene was not observed. These species which were observed in the case of the norbornane – benzene binary mixture were probably formed from cross-coupling reactions or from specific reactions of norbornane.

Mole fractions of biphenyl which were obtained in both studies (pure benzene and norbornane – benzene binary mixture) have been compared. The two graphs of Figure 11 display the evolutions of the mole fractions of biphenyl with residence time (Figure 11a) and with temperature (Figure 11b). This figure shows that the mole fraction of biphenyl is always slightly larger in the case of the norbornane – benzene binary mixture than in the case of benzene (apart from the experiment leading to the lowest conversion and performed with a temperature of 913 K and a residence time of 1 s; Figure 11b). It can also be observed that the variation between the mole fractions obtained during the two studies increases with conversion.

While interactions between the two hydrocarbons do exist during the thermal decomposition of the norbornane dissolved in benzene, the conditions of our study (temperatures less than 973 K) are such that the presence of benzene has a negligible influence on the reactions of norbornane. In a recent paper[26] El Balkali et al. showed that in the case of the oxidation of an equimolar n-heptane – benzene binary mixture the presence of benzene had very little influence on the reactivity at low temperature.

**Figure 11**

**Comparison with the reaction of pyrolysis of tricyclo[5.2.1.0$^{2,6}$]decane.** Experimental results obtained during this study have been compared with previous ones obtained with tricyclo[5.2.1.0$^{2,6}$]decane,[14] a tricyclic alkane the structure of which contains the structure of norbornane (Figure 12). This tricyclic alkane can be considered as a norbornane structure sharing two adjacent carbon atoms with a cyclopentane structure.

**Figure 12**

Figure 13 displays the evolution of the conversions of norbornane (dissolved in benzene) and tricyclo[5.2.1.0$^{2,6}$]decane obtained at temperatures ranging from 873 to 973 K, at a residence time of 1 s and at a pressure of 106 kPa. The mole percentages of norbornane and tricyclo[5.2.1.0$^{2,6}$]decane at the inlet of the reactor were set equal to 0.7%. This figure shows that the reactivities of the two polycyclic alkanes are very similar.

**Figure 13**

The thermal decomposition of tricyclo[5.2.1.0$^{2,6}$]decane leads to the formation of large amounts of hydrogen, ethylene, propene, 1,3-cyclopentadiene and cyclopentene.[14] While hydrogen, ethylene and 1,3-cyclopentadiene were also amongst the main products of the thermal decomposition of norbornane, propene appeared to be a minor product and the formation of cyclopentene was not observed. An analysis of the kinetic model of the pyrolysis of tricyclo[5.2.1.0$^{2,6}$]decane[14] shows that cyclopentene and allyl radicals (precursors of propene) mainly come from the cyclopentane part of the structure of tricyclo[5.2.1.0$^{2,6}$]decane.

Discussion

Most of the reactions involved in the pyrolysis of polycyclanes are still badly known and the related kinetic parameters are still very uncertain. We describe here the reactions involved in the mechanism of the thermal decomposition of the norbornane – benzene binary mixture and the possible channels of formation of the products of the reaction are discussed

**Unimolecular initiations by bond scission of norbornane. Fate of diradicals.** Unlike linear and branched alkanes for which two free radicals are directly obtained, unimolecular initiations of polycyclic alkanes by breaking of a C–C bond lead to the formation of diradicals (species with two radical centers). The molecule of norbornane (bicyclic alkane) has three different C–C bonds. The unimolecular initiations can lead to the formation of the three diradicals BR1, BR2 and BR3 shown on Figure 14. According to O'Neal and Benson[5] the activation energies of these reactions are given by the expression: $E_1=\Delta H(C–C)-\Delta E_{TC}+E_{-1}$, where E1 is the activation energy of the reaction of opening of the cycle, $\Delta H(C–C)$ is the bond energy of the broken C–C bond, $\Delta E_{TC}$ is the difference of ring strain energy between the products and the reactants, and $E_{-1}$ is the activation energy of the reverse reaction of closure of the diradical. If it is considered that the ring strain energy of norbornane[2] is equal to 17.2 kcal.mol$^{-1}$, that diradicals BR1 and BR2 have the same ring strain energy than cyclopentane (6,3 kcal.mol$^{-1}$) and that BR3 has the same ring strain energy than cyclohexane (0 kcal.mol$^{-1}$), the terms $\Delta E_{TC}$ are equal to 10.9 kcal.mol$^{-1}$ for BR1 and BR2 and to 17.2 kcal.mol$^{-1}$ for BR3. If we approximate that the sum $\Delta H(C–C)+E_{-1}$ is equal to about 87 kcal.mol$^{-1}$ (this was observed for cyclopropane, cyclobutane, cyclopentane and cyclohexane[14]) we obtain activation energies of 76.1 kcal.mol$^{-1}$ for BR1 and BR2 and of 69.8 kcal.mol$^{-1}$ for BR3.

**Figure 14**

In previous studies of the pyrolysis of cyclanes and polycyclanes[6,7,9,14] it has been shown that diradicals could react through three ways:

(1) by combination to give back the initial (poly)cyclane; this reaction is the reverse step of an unimolecular initiation by C–C bond scission.

(2) by internal disproportionnation through a (poly)cyclic transition state intermediate; an unsaturated molecule is then obtained.

(3) by decomposition by β–scission; products of the reaction depend on the position of the two radical centers. In most cases, a smaller diradical and a molecule are obtained.

**Figure 15**

Kinetic parameters of these reactions have been estimated for cyclobutane, cyclopentane and cyclohexane by quantum calculation by Sirjean et al..[9] This study showed that the easiest reaction is the reverse reaction by combination of the diradical formed by the unimolecular initiation (this is why cyclanes and polycyclanes present a greater stability than linear and branched alkanes). If we except this last reaction, the internal disproportionnation is largely easier than the decomposition by β–scission (apart from the particular case of cyclobutane in which the β–scission is easier because the broken C–C bond is in β position of the two radical centers). It is worth noticing that in the early stage of the reaction, (poly)cyclanes mainly lead to the formation of molecular species through diradicals and that they do not lead directly to the formation of free radicals. Thus at very low conversion the concentration of radicals is very low and the primary molecular initiation products from the reactant mainly react by unimolecular initiations to form new diradicals or free radicals.

Figure 15 displays the possible internal disproportionnations and decompositions by β–scission of diradicals BR1, BR2 and BR3 of Figure 14. For example BR2 can reacts by three different reactions of β–scission (to form two new diradicals of the same size and a smaller diradical with ethylene) and by three internal disproportionnations through bicyclic transition state intermediates (to form three unsaturated molecular species). New diradicals obtained by β–scission from BR1, BR2 and BR3 react in their turn by reactions of combination, disproportionnation and β–scission. Molecular species obtained through internal disproportionnation (which are the main products obtained from diradicals) react by unimolecular initiations to form diradicals and/or free radicals.

During experiments the formation of small amounts of 3-ethyl-cyclopentene, ethenyl-cyclopentane, 4-methyl-cyclohexene and methylene-cyclohexane (corresponding to MA4, MA2, MA6 and MA5 on the scheme of Figure 15) has been observed in the early stage of the reaction. These species can be obtained through internal disproportionnations of diradicals generated by the unimolecular initiations of norbornane and/or through metatheses of radicals involved in transfer and propagation reactions of the decomposition of the norbornyl radicals (but this last source of formation is in competition with more probable reactions of β–scission and isomerization). Thus 3-ethyl-cyclopentene, ethenyl-cyclopentane, 4-methyl-cyclohexene and methylene-cyclohexane likely come from the initiation step of norbornane (this is in accordance with the observation of the formation of these species at very low conversion). The formation of 1-methylene,3-methyl-cyclopentane (MA1 on Figure 15) has not been observed under the conditions of our study. This is probably because the diradical BR1 is more likely to react by reaction of termination by combination to give back norbornane. Unlike 3-ethyl-cyclopentene and ethenyl-cyclopentane the formation of 4-ethyl-cyclopentene (MA3 on Figure 15) was not observed. This may be explained by the fact that the bicyclic structure of the transition state which connects diradical BR2 and MA3 is more strained than the transition states which connect BR2 with MA2 and MA4. Study of the selectivity of the products of the reaction showed that MA2, MA4, MA5 and MA6 seemed to be primary products (extrapolations of their selectivities to origin gave values different from zero). Among these four species MA4 (3-ethyl-cyclopentene) has the highest selectivity at origin (Table 1) which also means that it has the highest initial rate of formation. This let us suppose that the unimolecular initiation of norbornane to diradical BR2 followed by the internal disproportionnation to MA4 is the easiest path of the initiation step.

Possible unimolecular initiations of 3-ethenyl-cyclopentene (main initiation product from norbornane) are given as an example (Figure 16). The breaking of the two vinylic C–C bonds of MA4 are not written on the scheme of Figure 16 because the bond dissociation energy is much higher than those of alkylic and allylic C–C bonds.

**Figure 16**

Reactions of unimolecular initiation by breaking of C–H bonds can also be considered but these reactions are more difficult than reactions of unimolecular initiation by breaking of C–C bonds. These reactions can lead to the formation of the three norbornyl radicals shown on Figure 17; but this channel of formation is negligible compared to the reactions of metathesis of radicals from norbornane (see below).

**Figure 17**

**Transfer and propagation reactions of norbornyl radicals.** The molecule of norbornane owns three different carbon atoms. Reactions of metathesis of hydrogen atoms and radicals with norbornane lead to the formation of three norbornyl radicals (Figure 18).

**Figure 18**

**Figure 19**

The three norbornyl radicals can react by decompositions by β–scission to lead to the formation of six cyclic radicals (Figure 19). These six new radicals can then react by decompositions by β–scission, by isomerizations and by metatheses (H abstractions) with molecules. Figure 20 displays the reactions of β–scission, metathesis and isomerization of radical R6 from Figure 19. Only isomerizations involving allylic hydrogen atoms have been taken in account on this scheme. The metatheses of radical R6 with molecules lead to the formation of 3-ethenyl-cyclopentane (MA4) but these reactions are less probable that the reactions of β–scission and isomerization.

Estimations of activation energies of reactions of β–scission by opening of the ring of cyclopentyl and cyclohexyl radicals[9,27] showed that the values used for linear and branched alkyl radicals[28-30] cannot be used in a systematic way for cycloalkyl radicals.[14] Norbornyl radicals have a more complex structure than cyclopentyl and cyclohexyl radicals and activation energies of their reactions of β–scission by opening of the ring are very uncertain. Moreover for some (poly)cyclic radicals (e.g. cyclopentyl radicals) the activation energy of the β–scission of C–C bonds may be much higher than the value used for linear and ramified alkanes[28-30] and the reactions of β–scission of C–H bonds may become competitive with the reactions of β–scission of C–C bonds.

**Figure 20**

Uncertainties on the kinetic parameters of reactions involved in the transfer and propagation steps of the norbornyl radicals make the discussion difficult at this stage of the study. Nevertheless it can be noticed that some radicals (like the radicals R6 and R7 of Figure 19) can lead to the formation of ethylene and cyclic C5 radicals which are precursors of 1,3-cyclopentadiene.

**Figure 21**

**Reactions of benzene.** Benzene is known for being a very stable hydrocarbon. At low temperature primary reactions of the pyrolysis of benzene are rather simple.[31] The only reaction of unimolecular initiation is the breaking of a C–H bond (bond energy of 110.9 kcal.mol$^{-1}$). Breaking of a benzylic C–C bond of benzene is very difficult because of the high energy of this type of bond (120.8 kcal.mol$^{-1}$). Bimolecular initiation (reverse reaction of a termination by disproportionnation) consists in the transfer of an hydrogen atom from a molecule of benzene to another. This step leads to the formation of a phenyl radical and a 2,4-cyclohexadien-1-yl radical. Activation energy of this reaction of bimolecular initiation (close to the enthalpy of the reaction: 94.4 kcal.mol$^{-1}$) is rather high. Under the conditions of our study, unimolecular and bimolecular initiations of benzene are probably negligible compared to unimolecular initiations of the norbornane which have lowest activation energies (part of the strain energy of norbornane is recovered during ring opening[3]).

The initiation step mainly leads to the formation of hydrogen atoms and phenyl radicals (Figure 21). Hydrogen atoms can reacts by reaction of metathesis with benzene to form an hydrogen molecule and a phenyl radical, by self combination to form hydrogen molecule (this trimolecular step is negligible) and by addition to benzene to lead to the 2,4-cyclohexadien-1-yl radical. Phenyl radicals can react by self combination or by addition to benzene (followed by the loss of an atom of hydrogen) to form biphenyl. Sivaramakrishnan et al.[32] observed the formation of acetylene and diacetylene from the decomposition by C–C bond β–scission of phenyl radical under extreme conditions (high temperature, high pressure benzene pyrolysis study behind reflected shock waves). 2,4-cyclohexadien-1-yl radical can react by C–H bond β–scission to give benzene, it can decompose by C–C bond β–scission to give 1,3,5-hexatrien-1-yl radicals and then acetylene and 1,3-butadien-1-yl radicals and can lead to 1,3-cyclohexadiene and 1,4-cyclohexadiene by metathesis on molecule. But under the conditions of the present study of benzene pyrolysis, i.e. at low temperature (below 973 K) and close to atmospheric pressure, decomposition of phenyl and 2,4-cyclohexadien-1-yl radicals does not occur and the formation of acetylene and unsaturated C4 hydrocarbons was not observed. This is in agreement with the observation of the formation of mainly hydrogen and biphenyl by Brooks and al. who studied the pyrolysis of benzene at temperatures between 873 and 1036 K in a static reactor.[25] According to Brioukov et al.[31], which performed the analysis of the experimental results obtained by Brooks and al.[25], the decomposition of benzene is dominated by the unimolecular initiation generating an hydrogen atom and a phenyl radical followed by the short propagation chain composed of the reaction of metathesis of an hydrogen atom with benzene and the reaction of ipso addition of a phenyl radical to benzene leading to biphenyl and an hydrogen atom.

**Cross-coupling reactions of the benzene – norbornane binary mixture.** Comparison between experimental results obtained during the study of the pyrolysis of pure benzene and that of norbornane dissolved in benzene showed that there are low interactions between the two hydrocarbons. There are very few possibilities of cross-coupling reactions (mainly reactions of metathesis and reactions of termination) because the primary mechanism of pyrolysis of benzene generates few species. Bimolecular initiations involving norbornane and benzene molecules lead to the formation of the three norbornyl radicals and of 2,4-cyclohexadien-1-yl radical (Figure 22). Activation energies of these reactions (between 80.5 and 89.4 kcal.mol$^{-1}$ according to the transferred hydrogen atom) are little higher than those of unimolecular initiation of norbornane but lower than the unimolecular initiation of benzene and than the bimolecular initiation of two molecules of benzene. Hydrogen atoms and radicals deriving from benzene (phenyl radicals and 2,4-cyclohexadien-1-yl radicals) can react by metatheses with norbornane to form the three norbornyl radicals (hydrogen, benzene, 1,3-cyclohexadiene and 1,4-cyclohexadiene are obtained respectively). In the same way radicals deriving from norbornane (more numerous than in the case of benzene) can react by metatheses with benzene to lead to phenyl radicals (Figure 22). Reactions of termination between radicals deriving from the two hydrocarbons can explain the formation of some products like toluene: this last species can be obtained from the combination between a phenyl radical (deriving from benzene) and a methyl radical (deriving from norbornane and not from benzene at such low temperature).

**Figure 22**

Conclusions

New experimental results of the thermal decomposition of norbornane dissolved in benzene have been obtained in a jet stirred reactor. A great attention was paid to the identification and the quantification of the products of the reaction. The formation of 25 both major and minor species has been observed during the experiments. Main products were hydrogen, ethylene and 1,3-cyclopentadiene. The detection of minor species having the same molecular weight as norbornane gave interesting information about the reactions of unimolecular initiation of norbornane. The study of the selectivities of the products of the reaction showed that 15 species were probably primary products and values of extrapolations to origin of selectivities let us think that the easiest initiation of the norbornane leads to the formation of 3-ethenyl-cyclopentene. The use of benzene as solvent of norbornane appeared to be a good choice because benzene was very unreactive under the operating conditions of our study: firstly interactions between the two hydrocarbons remained low and the reactivity of norbornane was little affected by the presence of benzene; secondly unimolecular initiations of norbornane (most probable initiation steps) were not masked by initiations involving benzene and the formation of molecular product from the diradicals generated through the initiations could be observed.

Reactions that occur during the pyrolysis of norbornane and during the pyrolysis of benzene have been reviewed and described. Cross coupling reactions in the case of the norbornane – benzene binary mixture are rather limited because the decomposition of benzene generates only few species at low temperature. The kinetics of the reactions of unimolecular initiation by breaking of the C-C bonds which are part of a ring structure, reactions of the diradicals (termination by combination, termination by disproportionnation and decomposition by β-scission) and reactions of decomposition by β-scission leading to the opening of a ring (e.g. reactions of β-scission of the norbornyl radicals) will require to be better investigated with reliable estimations of the kinetic parameters of these specific and sensitive reactions.

ACKNOWLEDGMENT

This work was supported by MBDA-France and the CNRS. We are grateful to E. Daniau, M. Bouchez, and F. Falempin for helpful discussion.

**Table 1.** Values of selectivities at origin of primary products of the reaction of pyrolysis of norbornane in benzene at a temperature of 953 K.

| Species | Selectivities to origin |
|---|---|
| hydrogen | 0.346 |
| methane | 0.011 |
| ethylene | 0.314 |
| ethane | 0.004 |
| propene | 0.020 |
| 1-butene | 0.006 |
| 1,3-butadiene | 0.035 |
| 1,3-cyclopentadiene | 0.184 |
| 1,3-cyclohexadiene | 0.013 |
| toluene | 0.004 |
| 3-ethyl-cyclopentene | 0.028 |
| 4-methyl-cyclohexene | 0.009 |
| methylene-cyclohexane | 0.007 |
| ethenyl-cyclopentane | 0.004 |
| biphenyl | 0.012 |
| *Total* | *0.997 (theoretical value: 1)* |

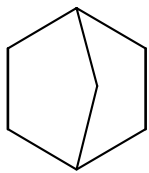

**Figure 1.** Structure of norbornane (bicyclo[2.2.1]heptane).

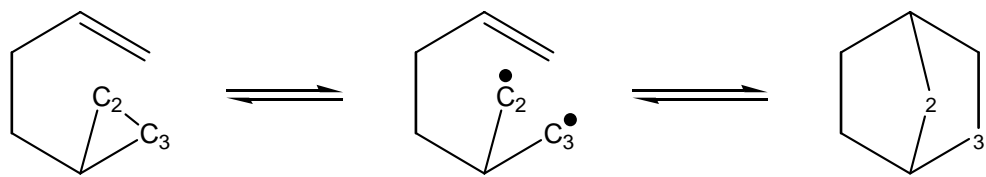

**Figure 2.** Isomerization of 3-butenyl-cyclopropane to norbornane through a diradical intermediate as proposed by Baldwin et al..[4]

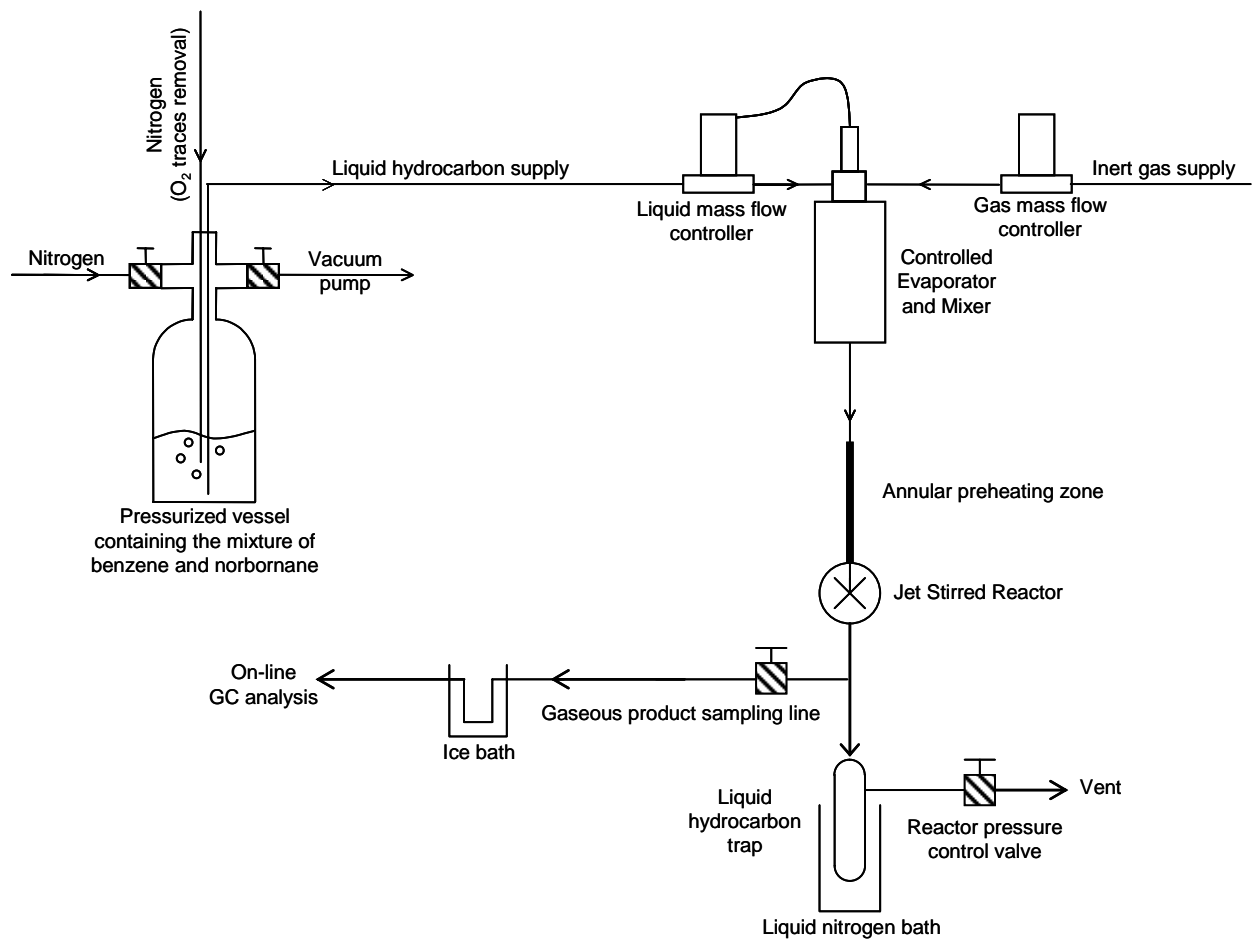

**Figure 3.** Experimental apparatus flow sheet.

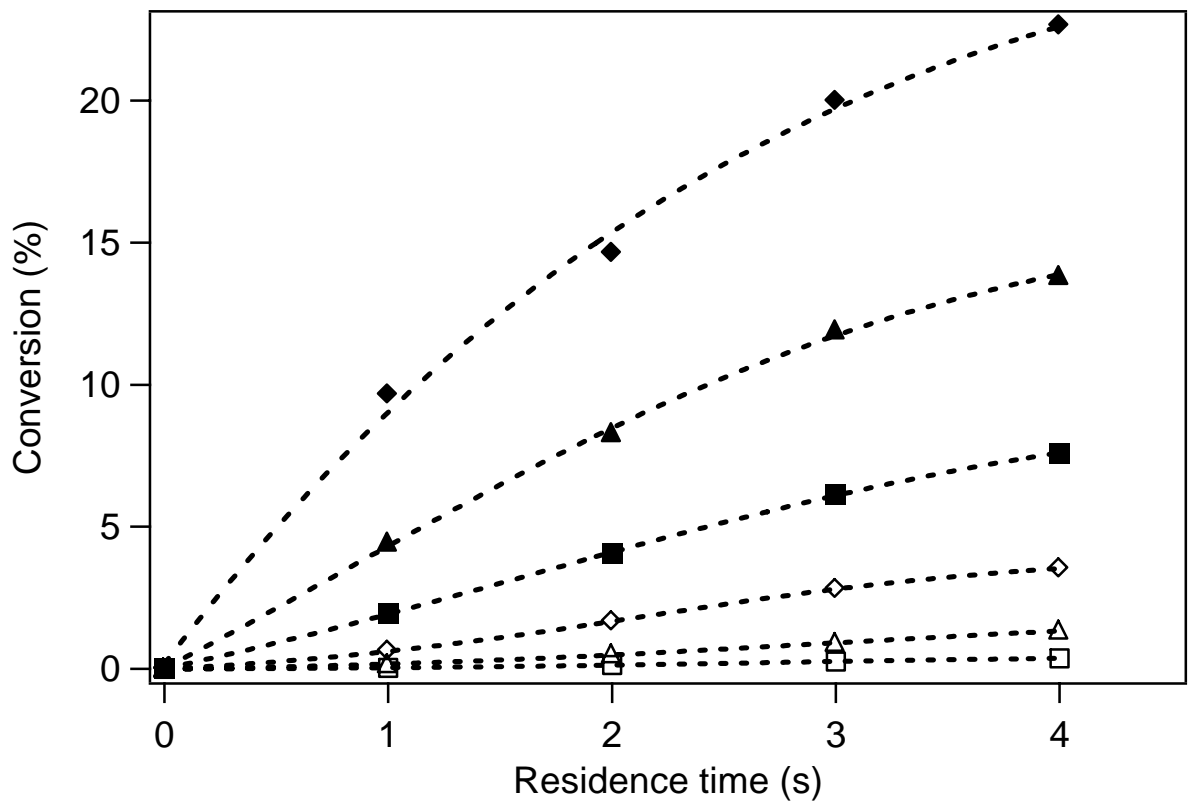

**Figure 4.** Evolution of the conversion of norbornane with residence time. (□ 873 K, △ 893 K, ◇ 913 K, ■ 933 K, ▲ 953 K, ◆ 973 K).

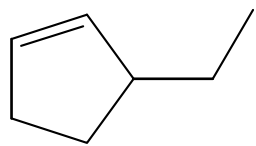 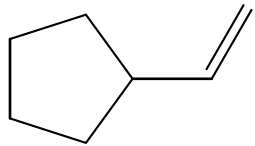 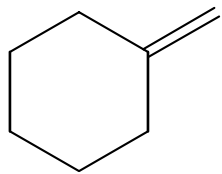 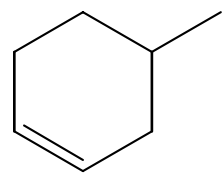

(a) (b) (c) (d)

**Figure 5.** Structure of products having the same molecular weight as norbornane. (a) 3-ethyl-cyclopentene, (b) ethenyl-cyclopentane, (c) methylene-cyclohexane and (d) 4-methyl-cyclohexene.

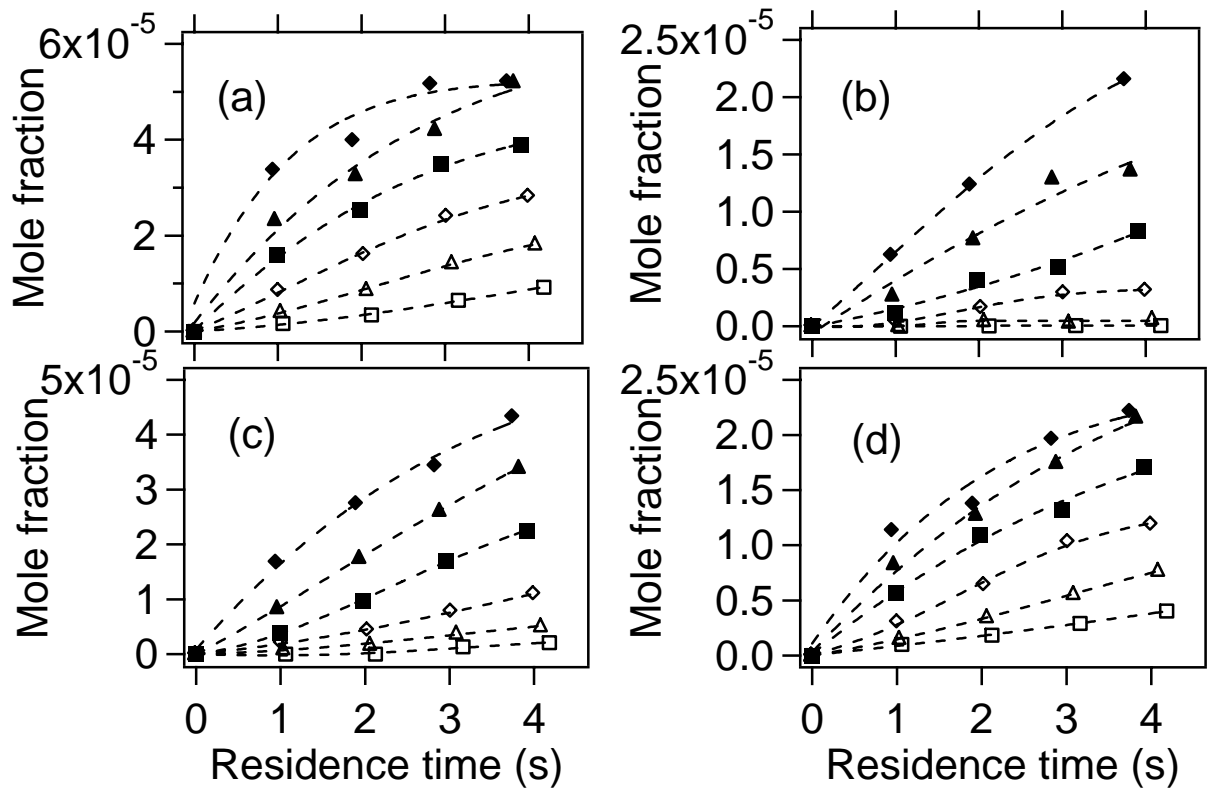

**Figure 6.** Evolution of the mole fractions of products having the same molecular weight as norbornane. (a) 3-ethyl-cyclopentene, (b) ethenyl-cyclopentane, (c) methylene-cyclohexane, (d) 4-methyl-cyclohexene. (□ 873 K, △ 893 K, ◇ 913 K, ■ 933 K, ▲ 953 K, ◆ 973 K).

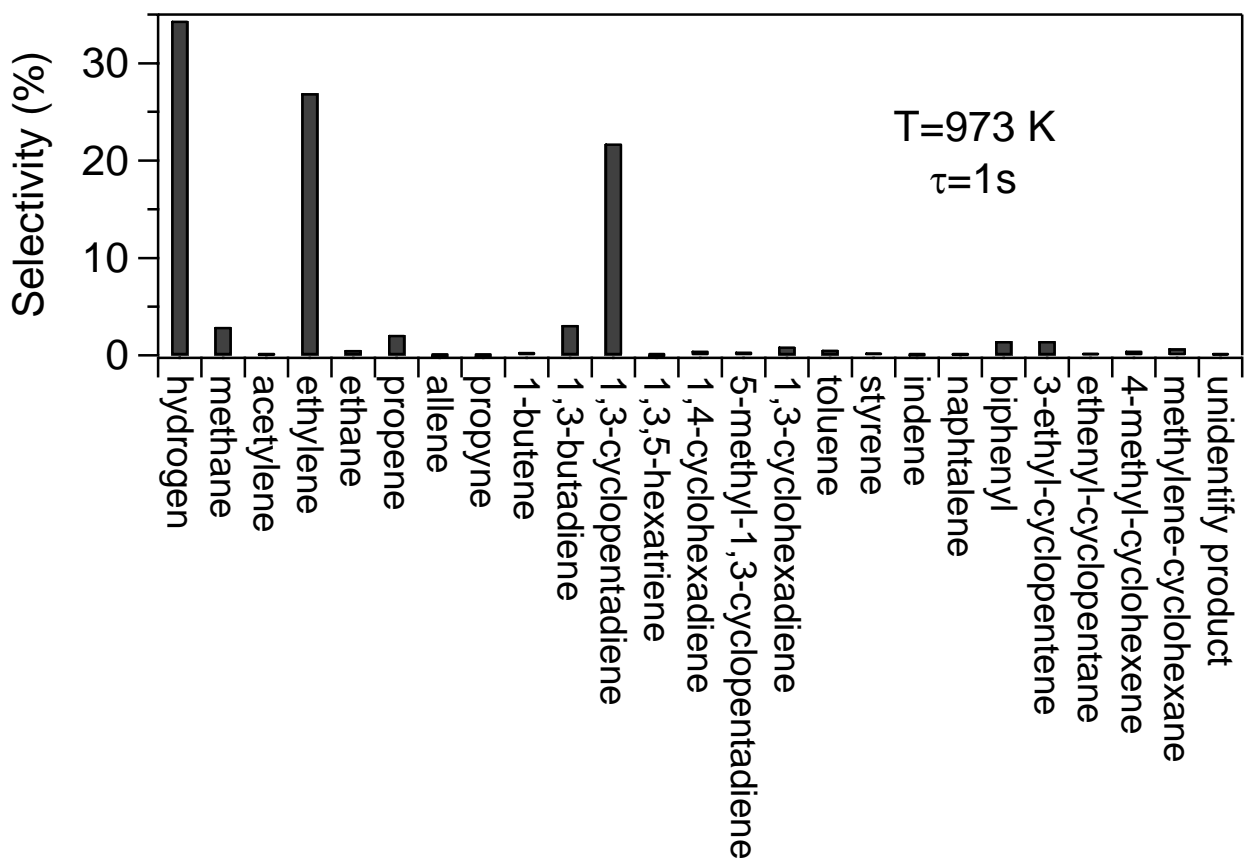

**Figure 7.** Distribution of the products of the reaction at a temperature of 973 K and at a residence time of 1 s.

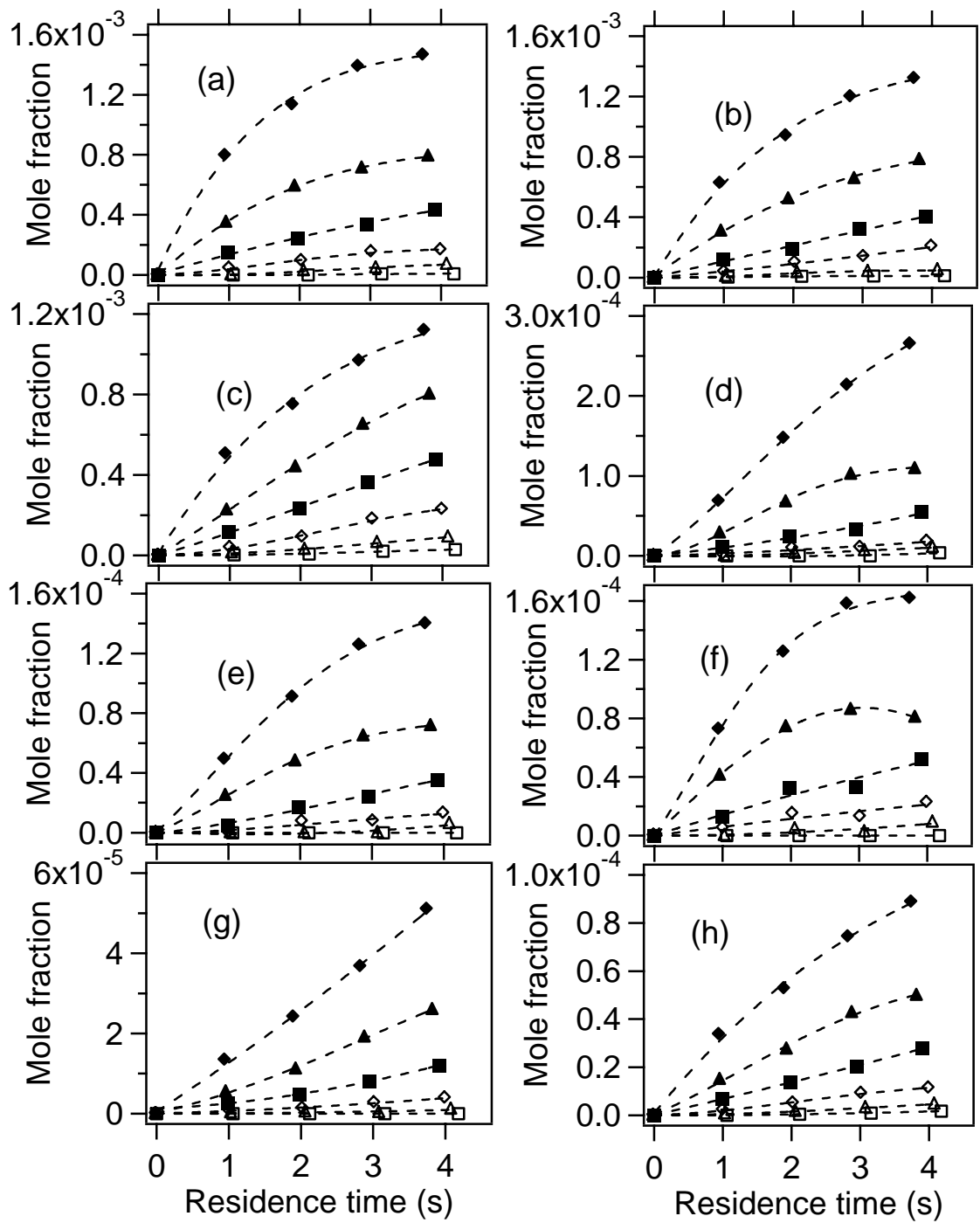

**Figure 8.** Evolution of the mole fractions of some products of the reaction with residence time. (a) hydrogen, (b) ethylene, (c) 1,3-cyclopentadiene, (d) methane, (e) propene, (f) 1,3-butadiene, (g) toluene, (h) biphenyl. (□ 873 K, △ 893 K, ◇ 913 K, ■ 933 K, ▲ 953 K, ◆ 973 K).
30

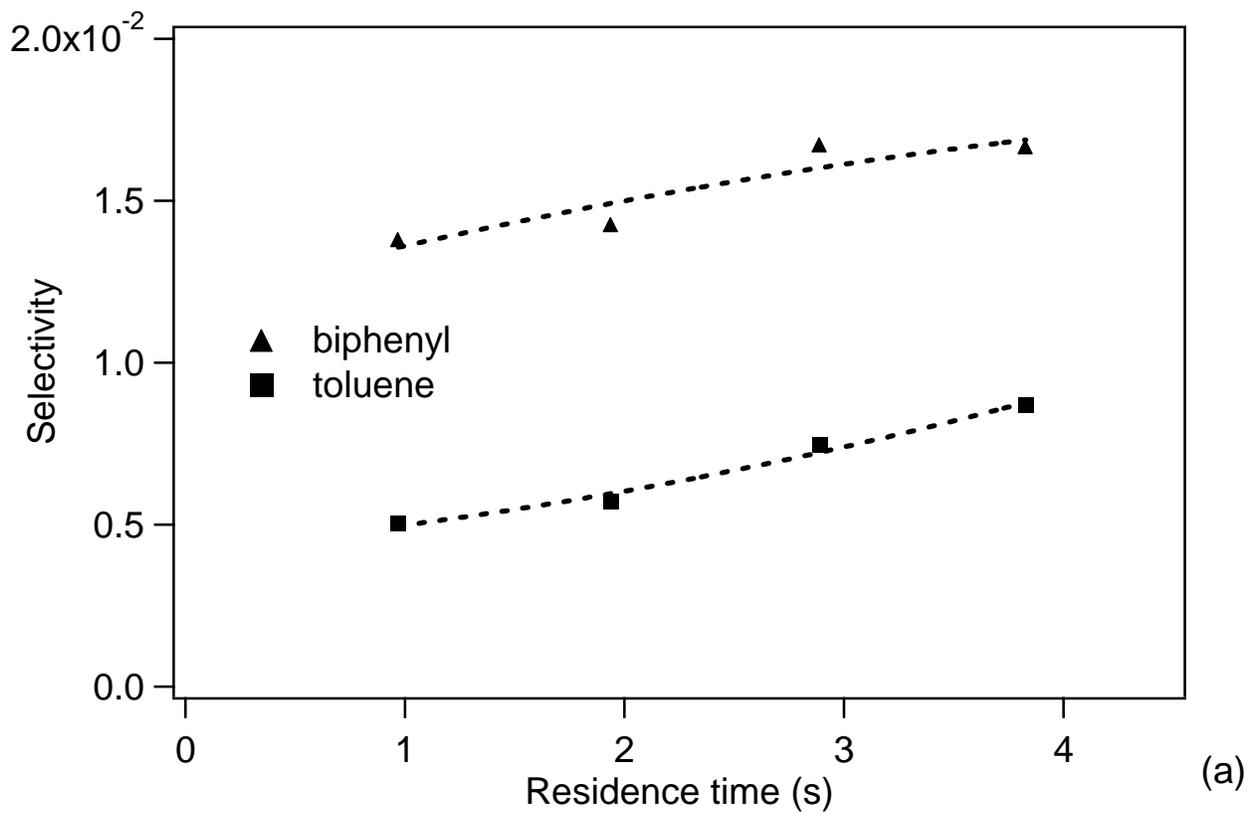

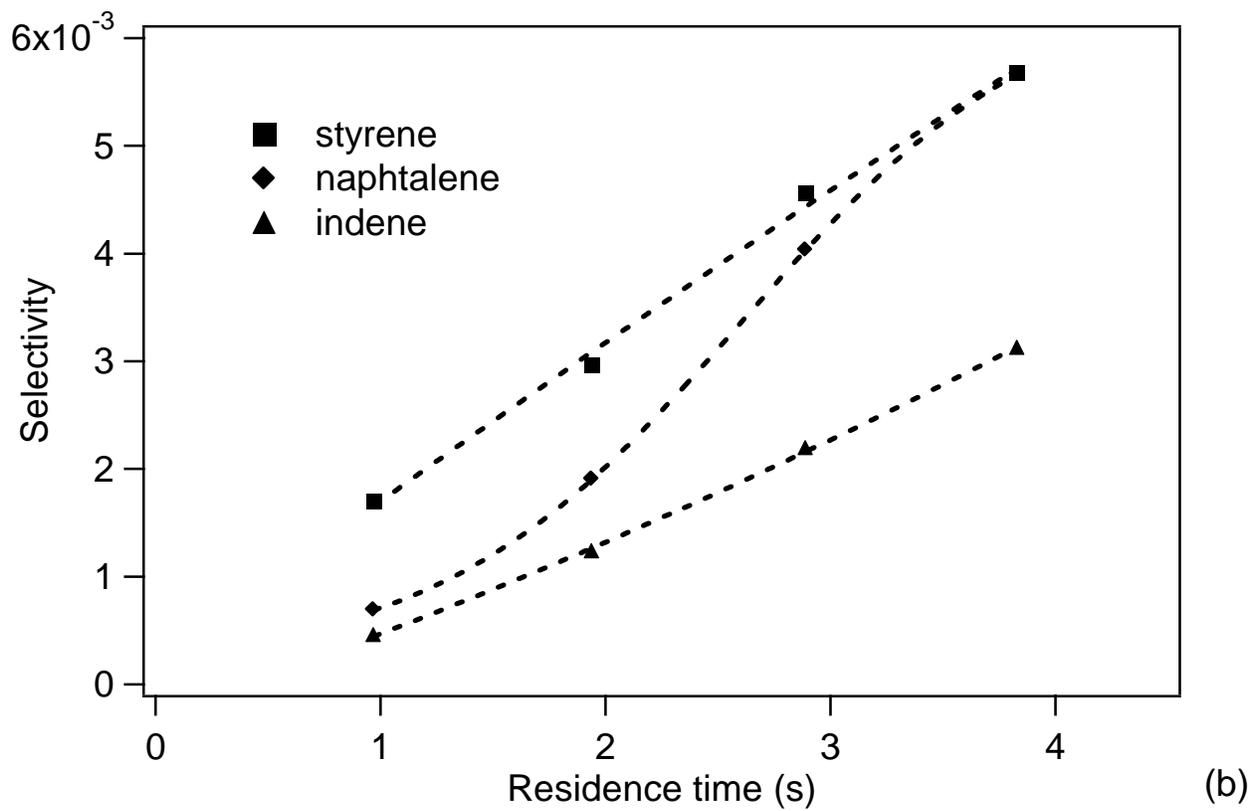

**Figure 9.** Evolution of the selectivities of (a) biphenyl and toluene and (b) styrene, indene and naphthalene with residence time (at a temperature of 953 K).

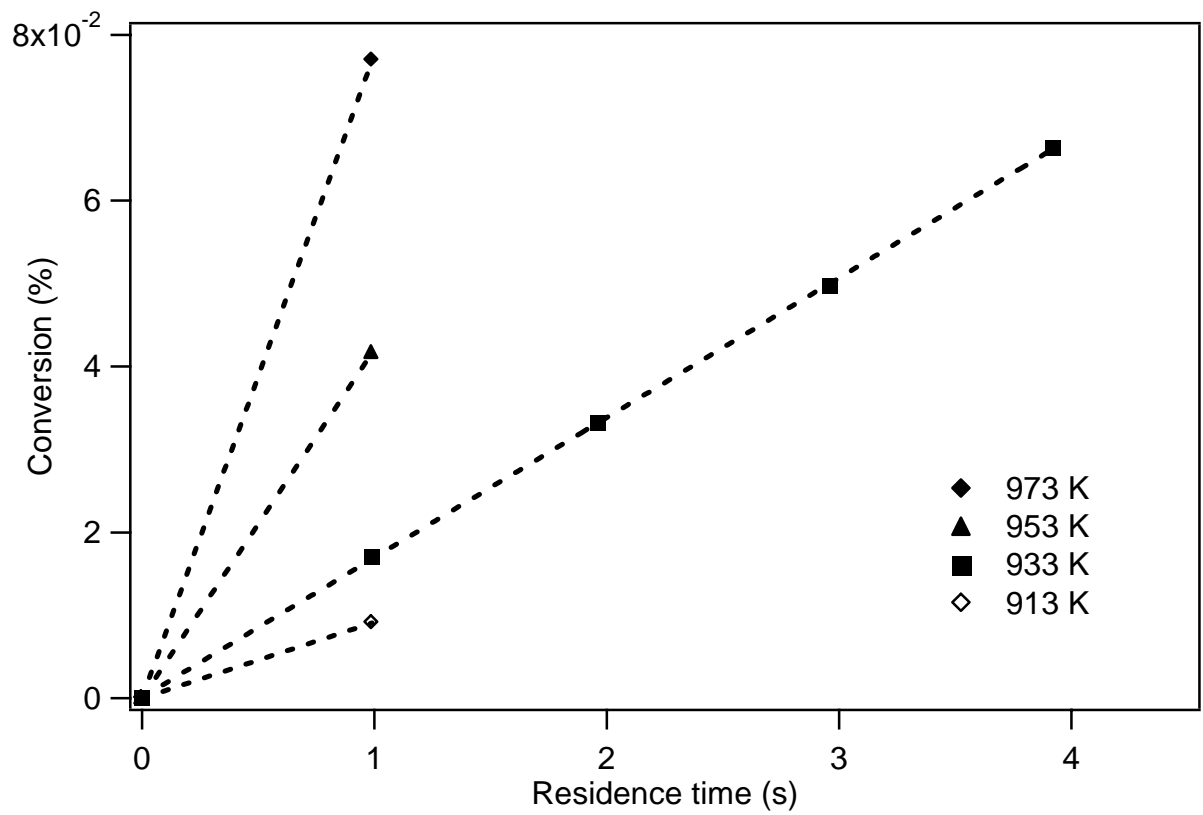

**Figure 10.** Evolution of the conversion of benzene with residence time.

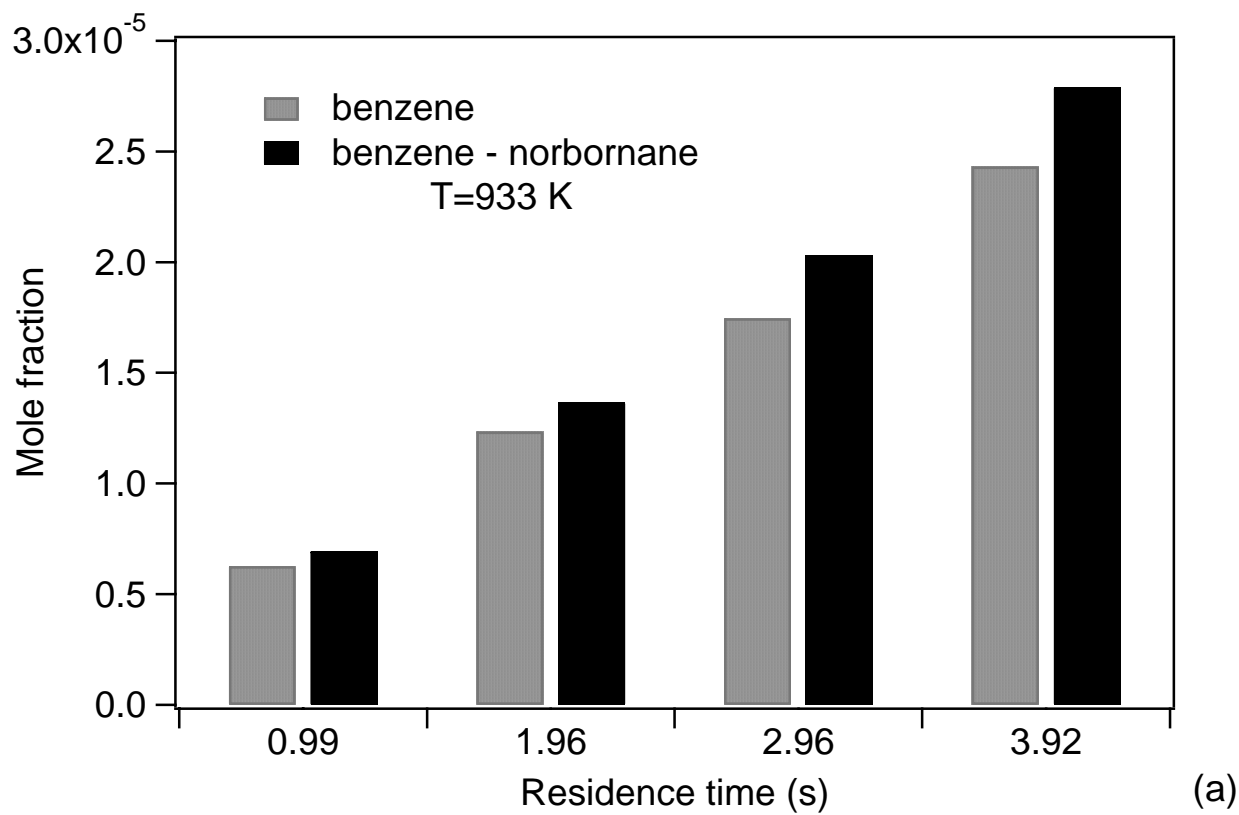

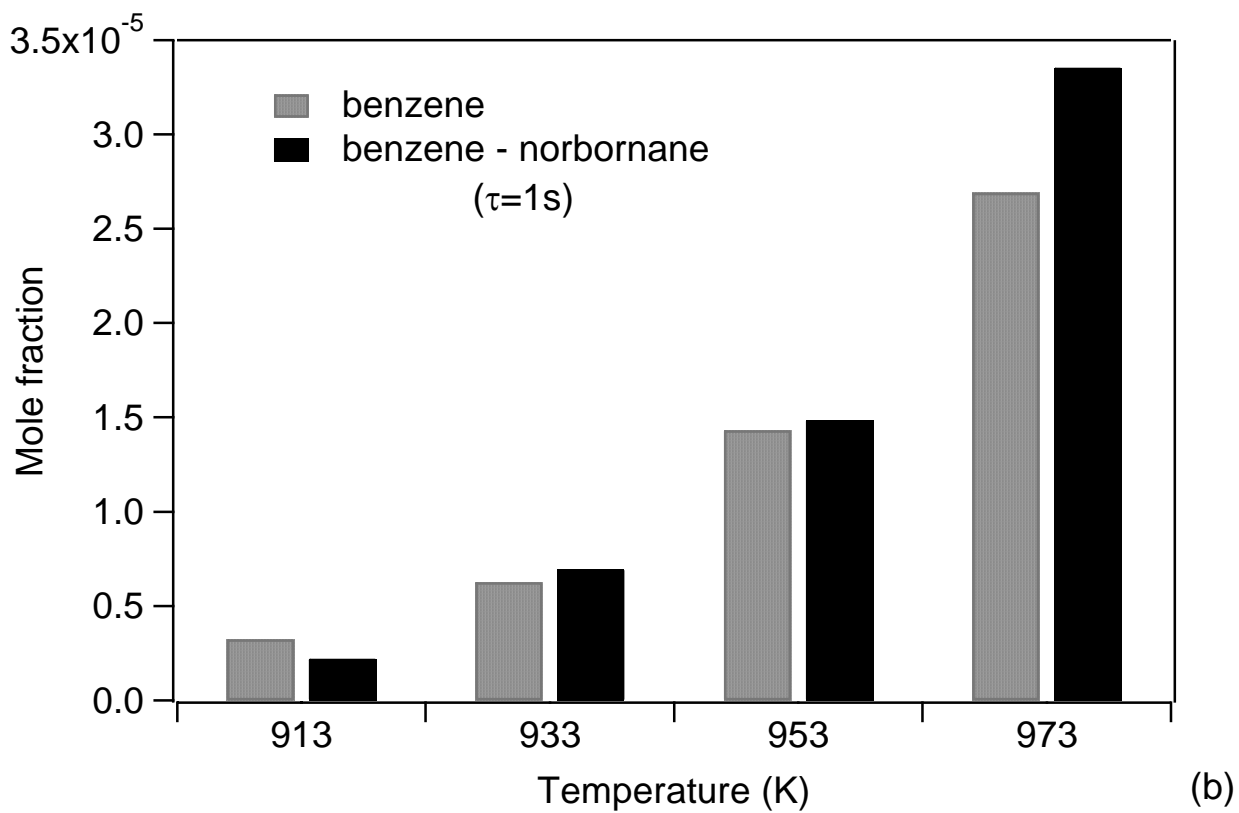

**Figure 11.** Comparison of mole fractions of biphenyl obtained in the two studies. Evolution with residence time (a) and with temperature (b).

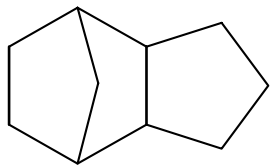

**Figure 12.** Structure of the tricyclo[5.2.1.0$^{2,6}$]decane.

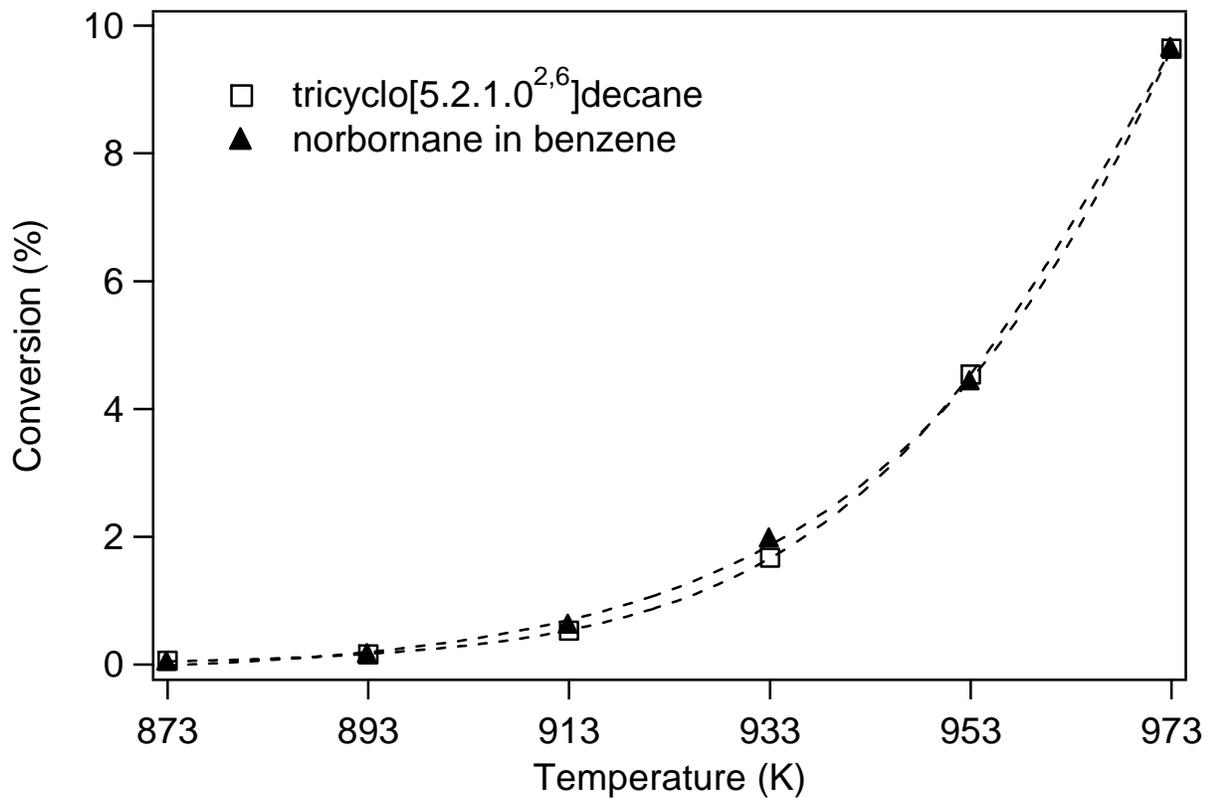

**Figure 13.** Comparison of the conversions of norbornane (dissolved in benzene) and tricyclo[5.2.1.0$^{2,6}$]decane. Experiments were performed at a residence time of 1 s, at a pressure of 106 kPa and with a mole percentage of reactant at the inlet of the reactor of 0.7%.

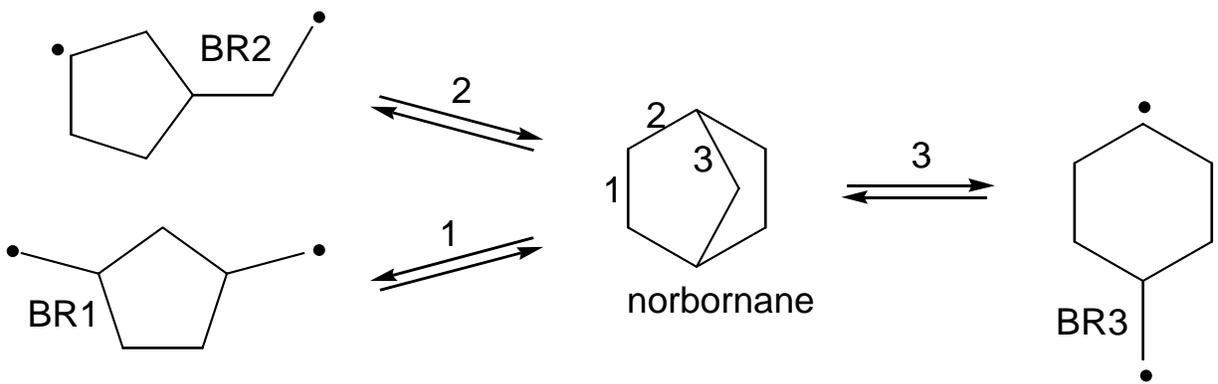

**Figure 14.** Reactions of unimolecular initiation of norbornane by breaking of C-C bonds.

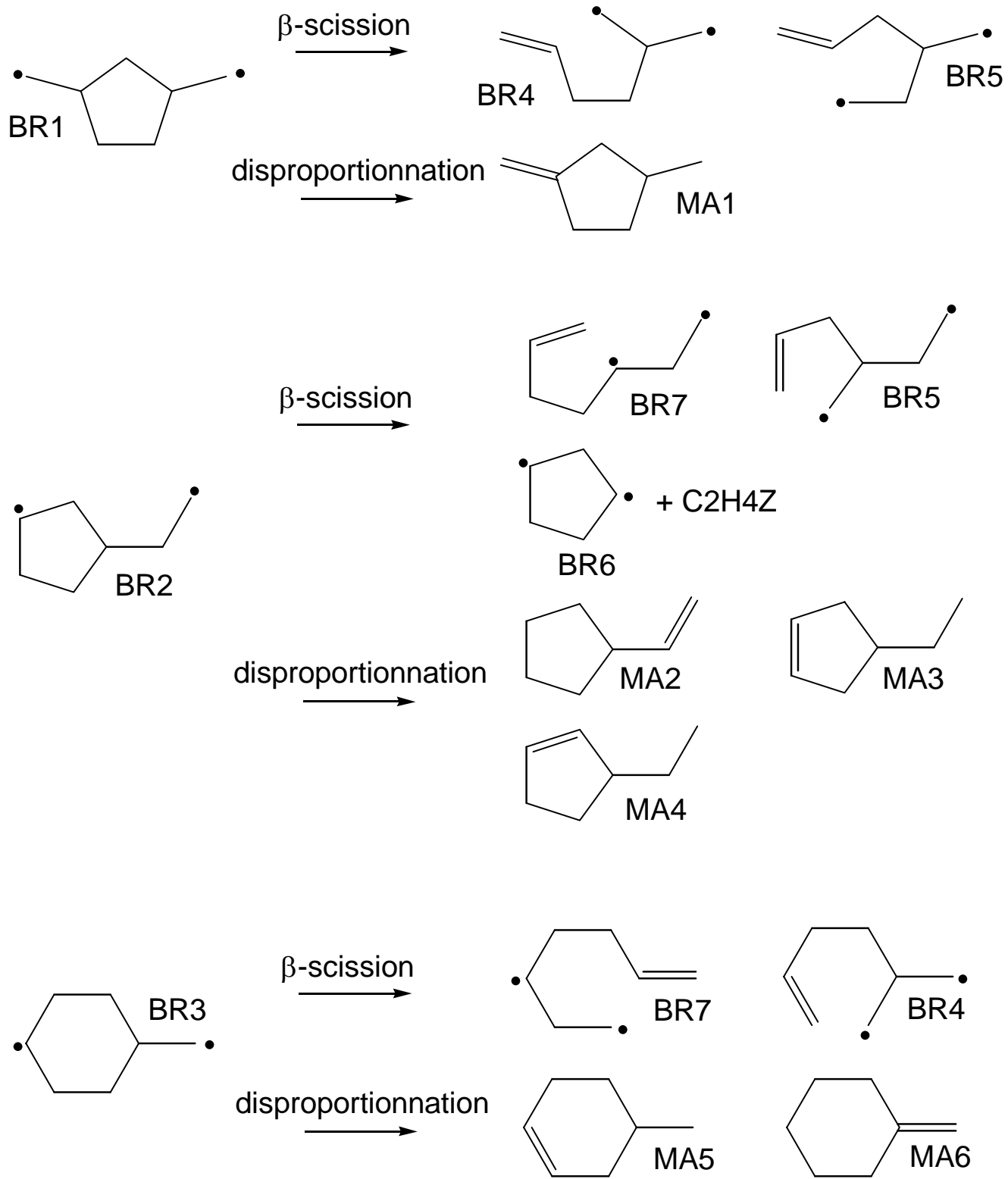

**Figure 15.** Reactions of β-scission and of disproportionnation of diradicals BR1, BR2 and BR3 of Figure 14.

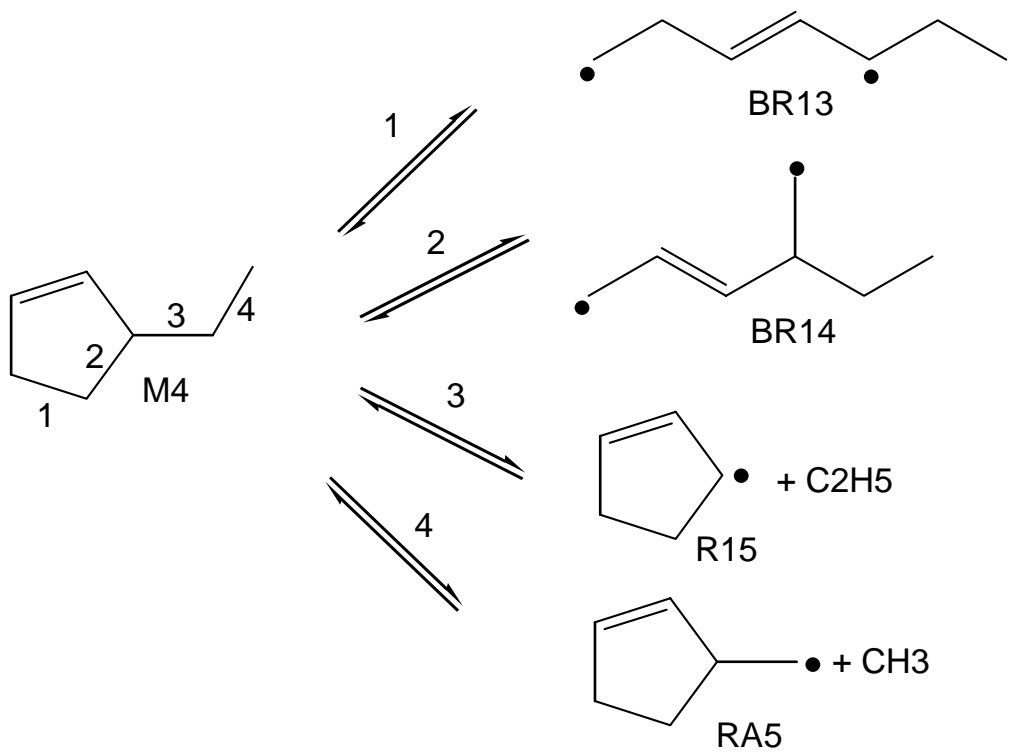

*Figure 16. Reactions of unimolecular initiation of 3-ethenyl-cyclopentene (MA4 of Figure 15).*

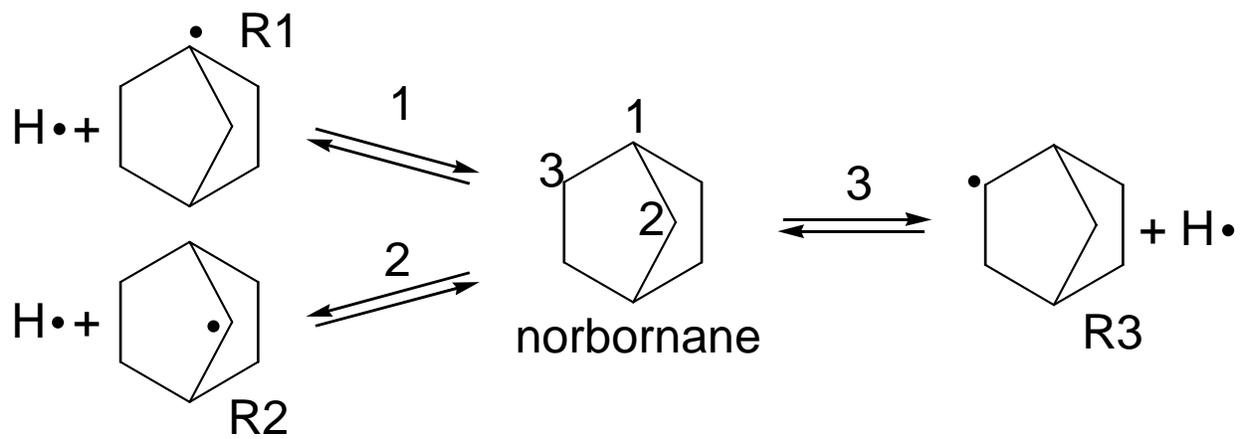

**Figure 17.** Reactions of unimolecular initiation of norbornane by breaking of C-H bonds.

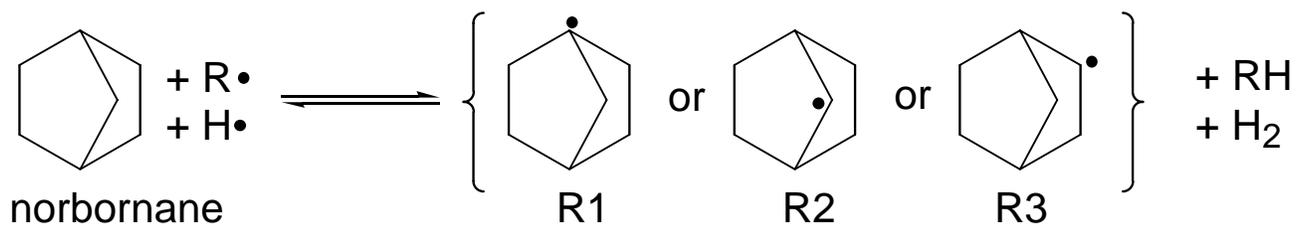

**Figure 18.** The three norbornyl radicals obtained by metatheses of hydrogen atoms or radicals R• on norbornane.

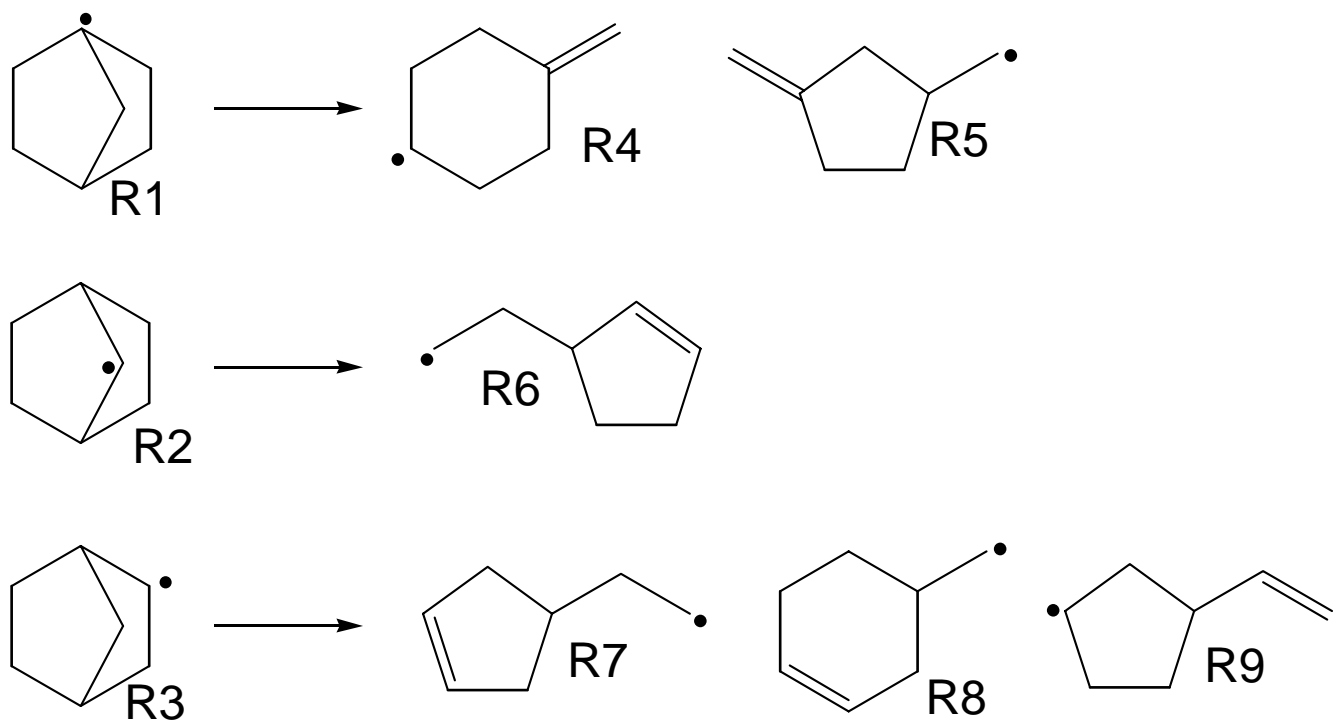
**Figure 19.** Reaction of decomposition by β-scission of the three norbornyl radicals.

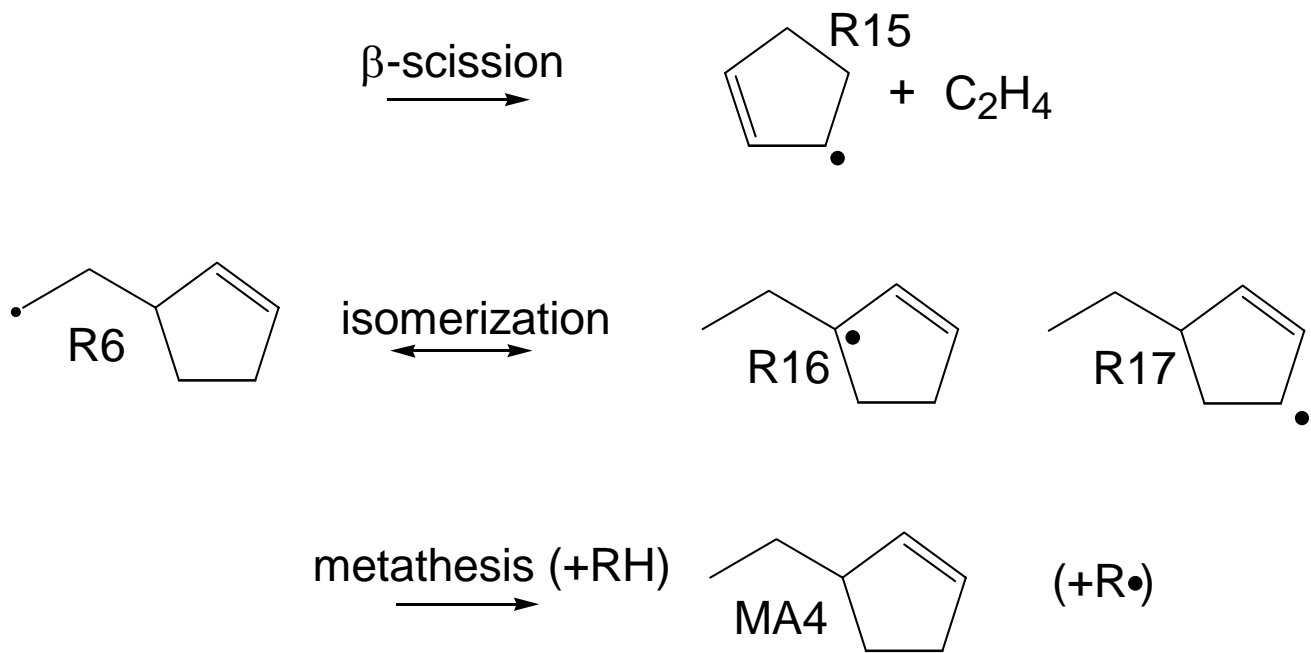

**Figure 20.** Reactions of the radical R6 of Figure 19.

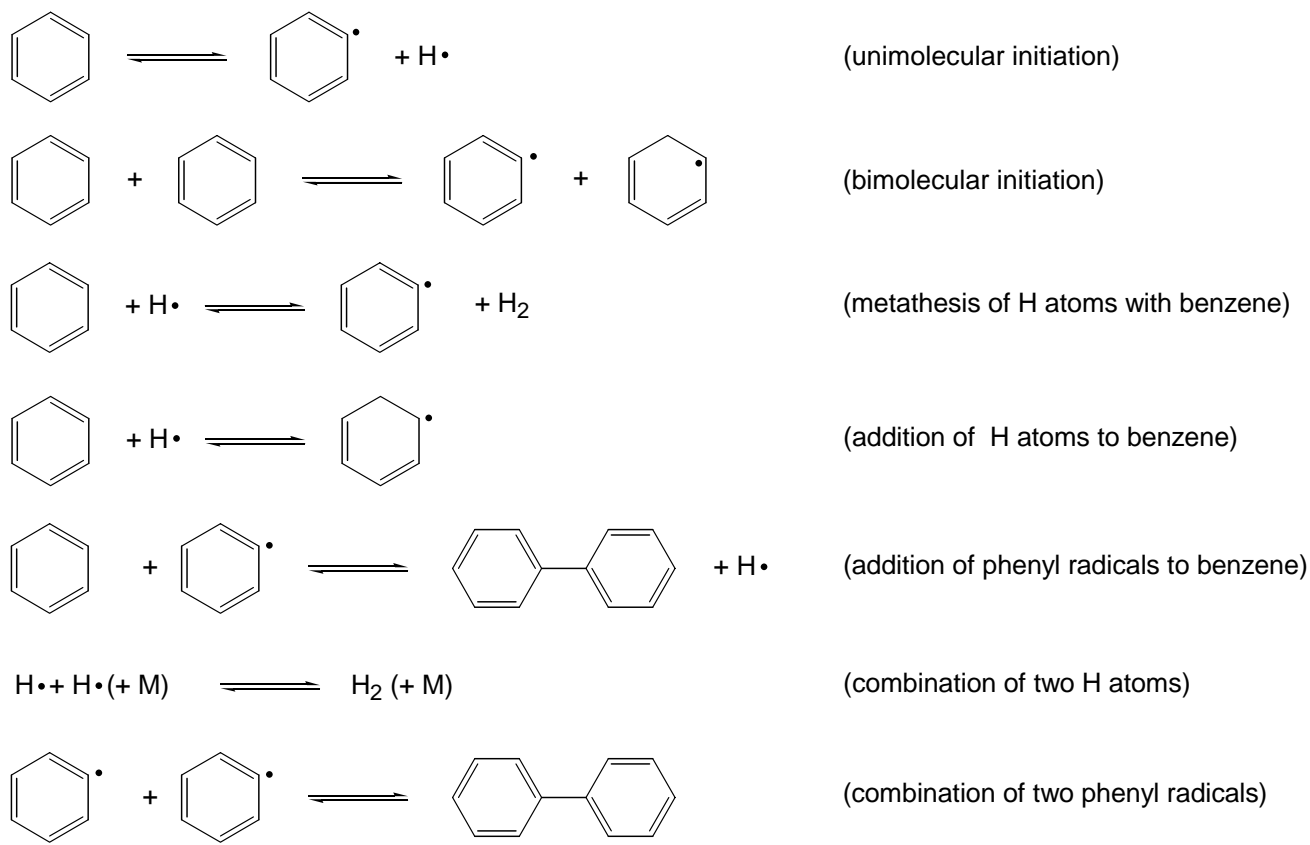

**Figure 21.** Primary reactions of the pyrolysis of benzene.

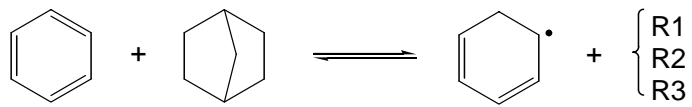

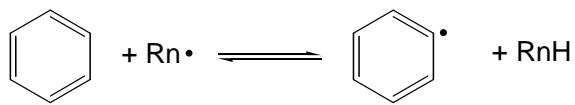

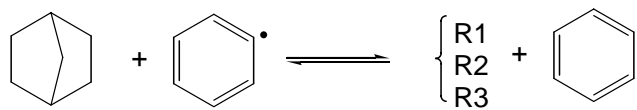

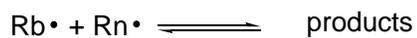

(bimolecular initiation)

(metatheses of radicals deriving from norbornane with benzene)

(metathesis of phenyl radical with norbornane)

(combination/disproportionnation of radicals deriving from norbornane and benzene)

**Figure 22.** Cross-coupling reactions of norbornane and benzene.